\documentclass[twocolumn]{bytedance_seed}

\usepackage[toc,page,header]{appendix}
\usepackage{microtype}
\usepackage{graphicx}
\usepackage{booktabs} 
\usepackage{pifont}
\usepackage{xcolor}

\usepackage{hyperref}

\newcommand{\sysname}{{\textsc{Comet}}}

\newcommand{\bluetext}[1]{\textcolor{black}{#1}}
\newcommand{\browntext}[1]{\textcolor{black}{#1}}

\usepackage{minitoc}

\title{\sysname{}: Fine-grained Computation-communication Overlapping for Mixture-of-Experts}

\author[1,2,\circ,*]{Shulai Zhang}
\author[1,\circ,\dagger]{Ningxin Zheng}
\author[1,\dagger]{Haibin Lin}
\author[1]{Ziheng Jiang}
\author[1]{Wenlei Bao}
\author[1]{Chengquan Jiang}
\author[1]{Qi Hou}
\author[2]{Weihao Cui}
\author[1]{Size Zheng}
\author[1]{Li-Wen Chang}
\author[2,\dagger]{Quan Chen}
\author[1,\dagger]{Xin Liu}

\affiliation[1]{ByteDance Seed}
\affiliation[2]{Shanghai Jiao Tong University}

\contribution[\circ]{Equal Contribution}
\contribution[*]{Work done at ByteDance Seed}
\contribution[\dagger]{Corresponding authors}

\abstract{
Mixture-of-experts (MoE) has been extensively employed to scale large language models to trillion-plus parameters while maintaining a fixed computational cost. 
The development of large MoE models in the distributed scenario encounters the problem of large communication overhead. The inter-device communication of a MoE layer can occupy $47\%$ time of the entire model execution with popular models and frameworks. Therefore, existing methods suggest the communication in a MoE layer to be pipelined with the computation for overlapping. However, these coarse grained overlapping schemes introduce a notable impairment of computational efficiency and the latency concealing is sub-optimal.

To this end, we present \sysname{}, an optimized MoE system with fine-grained communication-computation overlapping. Leveraging data dependency analysis and task rescheduling, \sysname{} achieves precise fine-grained overlapping of communication and computation. Through adaptive workload assignment, \sysname{} effectively eliminates fine-grained communication bottlenecks and enhances its adaptability across various scenarios.
Our evaluation shows that \sysname{} accelerates the execution of a single MoE layer by $1.96\times$ and for end-to-end execution, \sysname{} delivers a $1.71\times$ speedup on average. \sysname{} has been adopted in the production environment of clusters with ten-thousand-scale of GPUs, achieving savings of millions of GPU hours.
}

\correspondence{Ningxin Zheng, Haibin Lin, Quan Chen, Xin Liu}

\checkdata[Project Page]{\url{https://github.com/bytedance/flux}}

\begin{document}
\maketitle


\section{Introduction}
\label{sec:intro}

\bluetext{Recent advancements in large language models have revolutionized multiple domains, including natural language processing~\cite{vaswani2017attention, llama}, computer vision~\cite{liu2021swin} and multi-modal perception~\cite{liu2024visual, cao2023multi}. These achievements demonstrate that scaling up model size can significantly enhance model capacity.} However, the growth in model parameters poses substantial challenges for the deployment of such giant models,
as computational resources increasingly constrain model capacity~\cite{sharir2020cost}.

To this end, Mixture-of-Experts (MoE)~\cite{shazeer2017outrageously} introduces a sparse structure, within which only part of the parameters is activated. Instead of interacting with all parameters in dense models, MoE models allow each input to interact with only a few experts. 
For example, the Mixtral-8x7B model~\cite{jiang2024mixtral} comprises 45 billion parameters in total, while only 14 billion parameters are active during runtime. 
Nowadays, MoE has emerged as a key architecture for scaling models to trillion-plus parameters.

\begin{figure}[t]
\centering
	\includegraphics[width=\columnwidth]{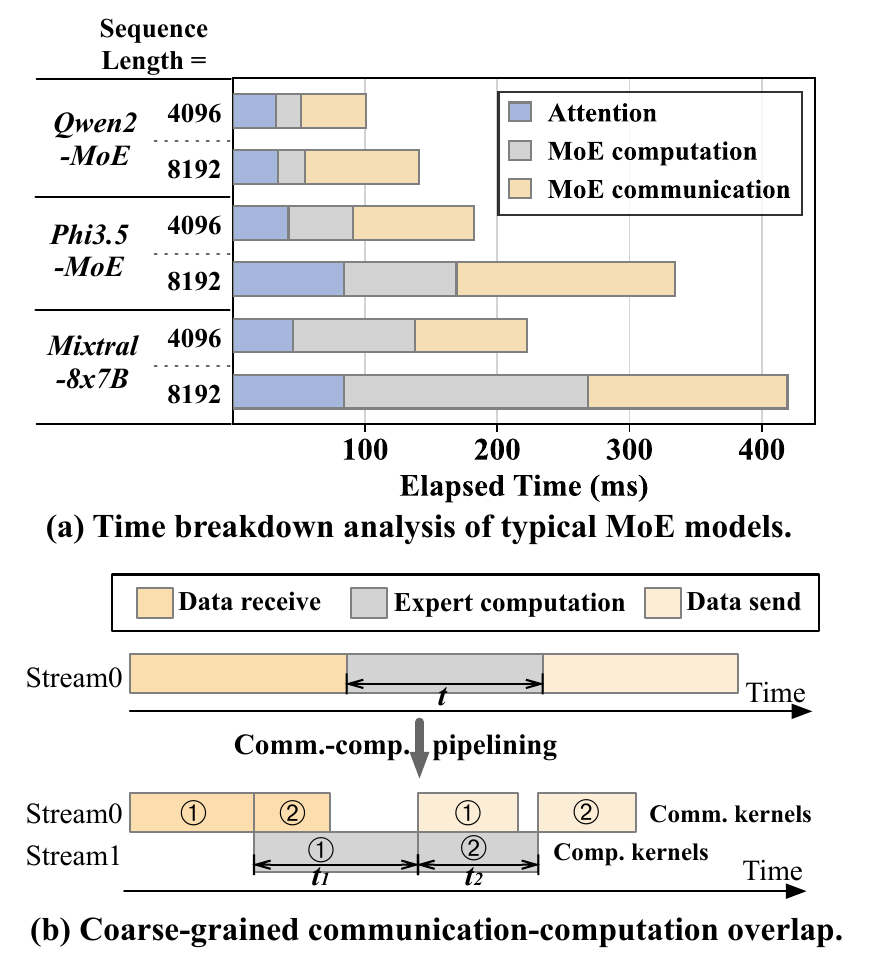}
	\caption{\label{fig:overall_breakdown} Analysis of the execution of MoE. (a) Time breakdown of MoE models executed on 8 H800 GPUs using Megatron-LM. \bluetext{(b) An illustration of communication-computation overlap by partitioning an expert computation kernel into two.}}

\end{figure}

The increase in parameter size in MoE models allows for the integration of greater amounts of information, but it poses challenges in expert placement. 
A typical approach is to distribute the experts across different GPUs as a single GPU cannot store all experts~\cite{gshard}. Consequently, during the execution of MoE layers, there is an intensive need for data exchange among GPUs.
In the forward pass of several popular MoE models, the communication among devices accounts for $47\%$ of the total execution time on average, as shown in~\autoref{fig:overall_breakdown}(a).


In a distributed environment, executing an MoE layer involves data reception, expert computation, and data transmission, as depicted in in~\autoref{fig:overall_breakdown}(b).
To reduce communication overhead, one effective strategy is to pipeline the process, overlapping communication with expert computation ~\cite{tutel, fastermoe, pipemoe, schemoe}. This approach involves partitioning input data into smaller data chunks, allowing decomposed communication and computation phases to overlap.
In the example in~\autoref{fig:overall_breakdown}(b), the received input data is divided into two chunks, and this coarse-grained overlapping reduces the overall execution time relative to non-pipelined execution.


\bluetext{The overlapping in existing mechanisms remains suboptimal due to two primary inefficiencies. 
First, the efficiency of partitioned experts declines as the data chunks assigned to each expert become smaller, potentially leading to under-utilization of GPU computational resources (e.g., the total compute time of experts after partitioning $t_1+t_2$ exceeds the original time $t$). 
The coarse-grained partitioning results in unavoidable GPU idle time during the initial and final communication phases, such as when receiving data for chunk 1 and sending data for chunk 2, which do not overlap with computation. Consequently, minimizing the non-overlapping time in these phases while maintaining computational efficiency is crucial. 
This is challenging because the data dependency between communication and computation is complex and it is hard to be overlapped in a fine-grained granularity efficiently. 
Second, due to the dynamic nature of MoE, the input shapes for experts are various at runtime, thereby posing diverse communication and computation burdens on GPUs. 
Encapsulating communication and computation tasks into separate kernels on different streams, like almost all the prior researches do, restricts control over hardware resources and results in non-deterministic kernel performance, thereby hindering seamless overlap (e.g., the computation of chunk 1 and the receiving of chunk 2 are misaligned). 
The second challenge, therefore, is to dynamically ensure precise allocation of hardware resources between computation and communication workloads at runtime.}
\bluetext{The complex data dependency, and the dynamic computation and communication workloads in MoE impede existing systems to realize efficient communication-computation overlap.} We therefore propose \sysname{}, a system that enables fine-grained communication-computation overlapping for efficient MoE execution. \sysname{} introduces two key designs: 
1) A dependency resolving method that identifies complex data dependencies between communication and computation operations in MoE, enabling optimized computation-communication pipeline structuring.
2) An adaptive workload assignment method that dynamically allocates GPU thread blocks to different workloads within a kernel, balancing communication and computation to improve latency concealment.


\sysname{} facilitates fine-grained overlapping in MoE by analyzing shared data buffers between communication and computation operations, referred to as \textit{shared tensor}. By decomposing the shared tensors along specific dimensions and reorganizing tensor data along with intra-operator execution order, \sysname{} eliminates the granularity mismatches between communication and computation, thereby enabling fine-grained overlapping.
To ensure precise resource allocation and effective latency concealment, \sysname{} integrates communication and computation tasks within fused GPU kernels. Through thread block specialization, 
\bluetext{\sysname{} isolates the impact of communication on computation performance}
, maintaining high computational efficiency. By adjusting the number of thread blocks allocated to each workload, \sysname{} effectively balances communication and computation latencies and reduces bubbles in overlapping.

We have integrated \sysname{} into Megatron-LM~\cite{megatron} and verified the capability of \sysname{} with various parallel strategies. Our extensive experiments on Nvidia H800 and L20 clusters show that \sysname{} delivers $1.96\times$ speedup for typical MoE layers, and $1.71\times$ speedup for end-to-end MoE model execution (Mixtral-8x7B~\cite{jiang2024mixtral}, Qwen2-MoE~\cite{bai2023qwen}, Phi3.5-MoE~\cite{abdin2024phi}) on average, compared with the SOTA MoE systems.  \bluetext{\sysname{} has been deployed to accelerate training and inference of large MoE models in production clusters comprising over ten thousand GPUs, achieving savings of millions of GPU hours.} 
\sysname{} introduces a fine-grained pipelined programming model for computation and communication. \textbf{We will open-source COMET, aiming to inspire further optimizations}, such as implementing the programming model in \sysname{} using compilers like Triton~\cite{triton} or TVM~\cite{chen2018tvm}. 

\section{Background and Motivation}


\subsection{MoE Structure}

\begin{figure}[t]
\centering
	\includegraphics[width=\columnwidth]{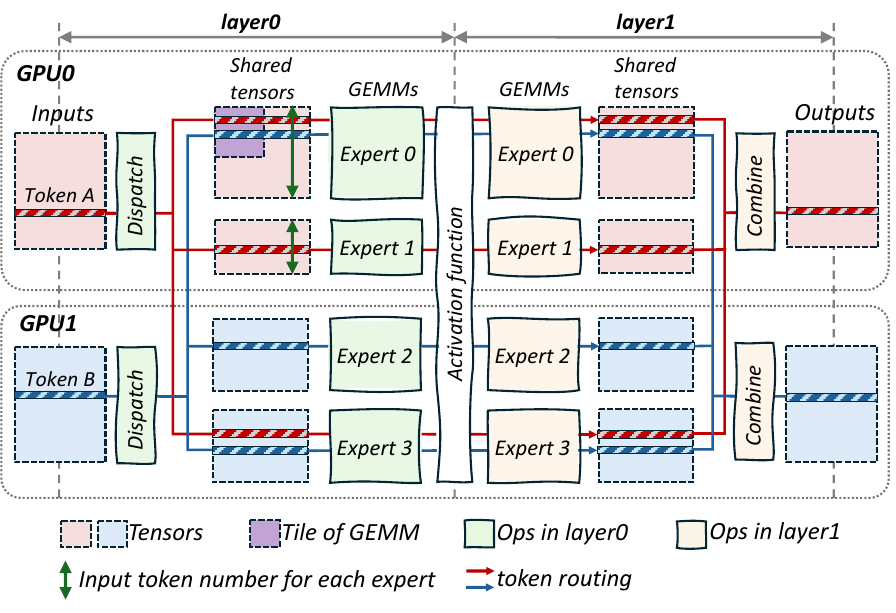}
	\caption{\label{fig:background}Example of an MoE layer across two GPUs, with two experts reside on GPU0 and two reside on GPU1. The MoE layer is composed of two feed-forward layers. In this example, for each token in the input buffer, it is dispatched to three experts ($topk=3$) in layer0 and then the results are combined in layer1. The shape of experts is $N\times K$ in layer0 and $K\times N$ in layer1.}
\end{figure}


\begin{table}[ht]
\centering
\footnotesize
\caption{\label{tab:description} Description of symbols.}
\begin{tabular}{@{}ll@{}}
\toprule
Symbol & Description \\ \midrule
$L$      & Number of transformer layers \\
$E$      & Total number of experts \\
$topk$   & Number of experts that each token is routed to \\
{\it TP}     & Tensor parallel size             \\
{\it EP}     & Expert parallel size             \\
$W$      & Total parallel world size ({\it TP}$\times$ {\it EP})  \\
$M$      & \bluetext{ Input token length  $\times$ Batch size }           \\
$N$      & Embedding size of a token            \\
$K$      & Hidden size of the feed-forward layer in experts            \\
\bottomrule
\end{tabular}
\end{table}

Mixture of Experts (MoE) is critical for efficiently scaling models. By enabling sparse activation of parameters, MoE allows for the integration of more parameters without increasing execution costs, thereby enhancing performance. 
The key idea of MoE is that it consists of multiple small models, namely \textit{experts} and tokens are only routed to partial experts for computation. \autoref{fig:background} shows the typical execution flow of an MoE layer and \autoref{tab:description} explains symbols to describe the execution of an MoE model.


Each input token is assigned to one or more experts for computation, with assignments determined by various algorithms~\cite{stochastic, zhou2022mixture, liu2022gating}. A common method involves a gate network ~\cite{shazeer2017outrageously} that selects the $topk$ experts for each token, as shown in~\autoref{fig:background}, where token A is routed to Expert0, Expert1 and Expert3.
After passing through two feed-forward layers of General Matrix Multiply (GEMM), the $topk$ outputs are gathered and reduced to produce the final result.


The operations in MoE’s layer0 comprise token communication (dispatch) across GPUs and the first layer of expert computations (GEMM operations), thereby establishing a communication-computation pipeline. MoE’s layer1 includes the second layer of expert computations, \bluetext{token undispatch and the topk reduction (combine)}, forming a computation-communication pipeline.

MoE employs two primary parallelization strategies: \textbf{Expert parallelism}~\cite{gshard} and \textbf{Tensor parallelism}~\cite{megatron}. In expert parallelism, the weights of different experts are distributed across separate GPUs, with each expert’s weights being fully intact. Tokens are routed to the corresponding devices of their respective experts. \autoref{fig:background} shows a case for expert parallelism, with Expert0 and Expert1 reside on GPU0 and others reside on GPU1. In contrast, tensor parallelism partitions the weights of all experts along the hidden dimension, with each GPU hosting a portion of the weights from all experts. Both expert and tensor parallelism are essential for the efficient execution of MoE. In practical deployment of MoE models, a hybrid parallelism approach combining both expert and tensor parallelism is often applied.

\subsection{Computation and Communication Overlapping}

As the MoE architecture grows larger and sparser, the proportion of time spent on communication in MoE models becomes increasingly significant, as shown in~\autoref{fig:overall_breakdown}(a). As illustrated in \autoref{sec:intro}, coarse-grained overlapping of computation and communication offers limited optimization potential, and \browntext{kernel-level scheduling is not efficient for dynamic workloads. Thus, it is more efficient to perform the overlapping at a fine-grained granularity (such as token-wise) and integrates computation and communication workloads into fused GPU kernels.}
Adopting such a finer-grained overlapping could extremely unleash further optimization opportunities. 
\bluetext{However, achieving such fine-grained overlapping in MoE is non-trivial and there are two primary obstacles in our observation.}

\begin{figure*}[t]
\centering
	\includegraphics[width=1.7\columnwidth]{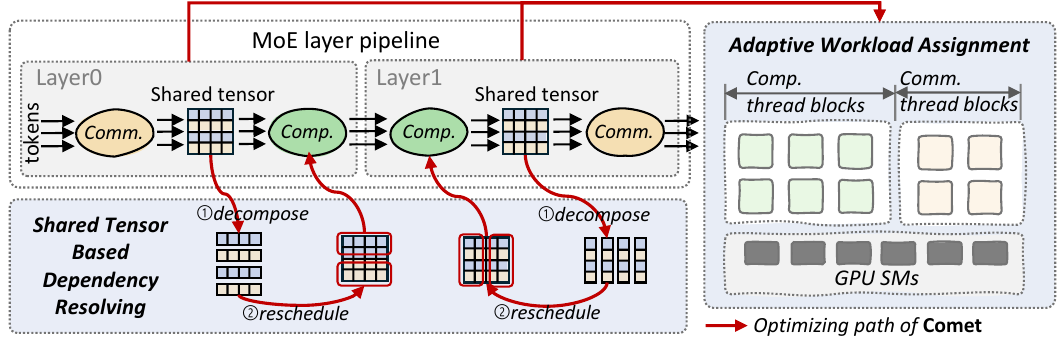}
	\caption{\label{fig:overview} Design overview of \sysname{}. \browntext{\sysname{} is composed of a shared tensor-based dependency resolving method and an adaptive workload assignment mechanism.}}
\end{figure*}

\subsubsection{Granularity mismatch between computation and communication}
In MoE systems, the token serves as the fundamental unit of data movement, illustrated by the movement of Token A in \autoref{fig:background}. To maximize GPU compute efficiency, high-performance GEMM(GroupGEMM) kernels typically organize rows into tiles for processing. The purple block in \autoref{fig:background} represents such a computation tile in GEMM kernels, exemplified by a 128x128 tile. Therefore, the GEMM computations associated with a single expert may require 128 tokens distributed across multiple GPUs. When fusing computation and communication at fine granularity, the disparity between token-level data transfer and tile-level computation introduces considerable challenges\browntext{: The complex data dependency adversely affects the efficiency of overlap, prompting the use of fine-grained communication, while integrating fine-grained communication with computation within fused kernels is also challenging.}

\textbf{Complex data dependency.}
\bluetext{The tokens needed for each computation tile, determined by the MoE’s gate at runtime, are randomly distributed across multiple devices. Computation for a tile cannot start until all required tokens are available.}
As shown in~\autoref{fig:background}, Expert0’s tile does not initiate processing until both Token A and Token B are received. Thus, with coarse-grained data communication, data preparation time for each computational tile may be prolonged because of this irregular and complicated data dependency. To mitigate this, we should employ fine-grained communication, where each computational tile reads or writes only the data it requires \bluetext{directly through the Unified Virtual Address~\cite{uva}}, and leverage the data reorganization and rescheduling to hide it with computation efficiently.

\textbf{Fine-grained communication.}
The integration of token-wise communication with tile-wise computation for overlapping is non-trivial. 
\bluetext{Remote I/O operations between GPUs exhibit significantly higher latency compared to local GPU memory access. Therefore, executing numerous fine-grained read and write operations on remote data tokens within computation thread blocks can block subsequent computational tasks, leading to a significant decline in kernel efficiency. This challenge is especially evident in the Hopper architecture, where computation kernels leverage Tensor Memory Accelerator (TMA) hardware instructions~\cite{tma} to establish asynchronous compute pipelines. The integration of long-latency remote I/O operations within these asynchronous pipelines can considerably prolong the overall execution time, adversely affecting performance.
Thus, it is critical to constrain the impact of fine-grained communication on computation kernels. }

Our first insight is that resolving the granularity mismatch between computation and communication in MoE models is the key to enable efficient overlap of these two processes.

\subsubsection{Diverse loads of computation and communication}
\browntext{Another characteristic of MoE is the dynamic routing of tokens to different experts, resulting in varying input shapes for experts at runtime (e.g., the token number received by Expert0 and Expert1 are different as shown in~\autoref{fig:background}).
This variability imposes differing communication and computation demands on GPUs.}
Besides, the hardware environments can also have various compute architectures or network topologies, providing different compute capacities and communication bandwidths. 
Achieving seamless overlap between computation and communication thus requires dynamically adjusting the allocation of GPU resources to different workloads, which is hard to be realized through wrapping workloads into separate kernels.

Our second insight is that the resource allocation should be adaptive within kernels at runtime to further achieve seamless communication-computation overlapping.

\section{Design of \sysname{}}

In this section, we present the core design of \sysname{}, a Mixture of Experts (MoE) system optimized for efficient execution of MoE layers through pipelined execution and fine-grained overlapping of communication and computation.
Our analysis reveals that the MoE architecture has two distinct producer-consumer pipelines: the communication-computation pipeline and the computation-communication pipeline, as illustrated in~\autoref{fig:overview}.
Tokens traverse the pipelines as depicted and the operations within each pipeline are linked through a shared buffer, referred to as the \textbf{shared tensor}, serving as both the producer’s output buffer and the consumer’s input buffer.
To minimize overall latency and enhance pipeline performance, \sysname{} introduces two key mechanisms aimed at overlapping computation and communication workloads effectively.

1. \bluetext{Shared tensor based dependency resolving}: 
As previously mentioned, the intricate data dependencies between communication and computation pose a challenge to achieving seamless overlap between these operations. To address this, we examine the data dependencies by analyzing the shared tensor. Our analysis reveals that the shared tensor can be decomposed, and the associated computations can be rescheduled to overlap more effectively with communication. Accordingly, \bluetext{the dependency resolving process} employs two key optimization strategies on the shared tensors as shown in~\autoref{fig:overview}: \ding{172} Decomposing the shared tensors along specific dimensions to break the coarse-grained data dependencies and, \ding{173} rescheduling the computations to enhance efficiency while ensuring effective overlapping. 




2. Adaptive workload assignment:
Following pipeline optimization by the dependency \bluetext{resolving}, the pattern of communication-computation overlap becomes more consistent and regular. To effectively hide the fine-grained communication latency, it is essential to allocate appropriate hardware resources to both communication and computation workloads. Given that these workloads exhibit different performance characteristics depending on input shapes, model configurations, and hardware environments, the adaptive workload assignment scheme dynamically balances computation and communication. This approach generates highly efficient horizontally-fused kernels for the MoE system, thereby optimizing latency concealment.

\browntext{As shown in~\autoref{fig:overview}, \sysname{} first leverages the shared tensor based dependency resolving method to optimize the pipelines in the MoE structure by decomposing and rescheduling the shared tensors. According to the reformed pipelines, \sysname{} then provides highly-efficient fused kernels through the adaptive workload assignment mechanism.}

\subsection{Shared Tensor Based Dependency Resolving}
\bluetext{We now introduce how to resolve the complex data dependency between computation and communication in MoE}.
It aims to bridge the granularity of communication and computation operations to sustain high efficiency by decomposing and rescheduling shared tensors.

\begin{figure}[t]
\centering
	\includegraphics[width=0.95\columnwidth]{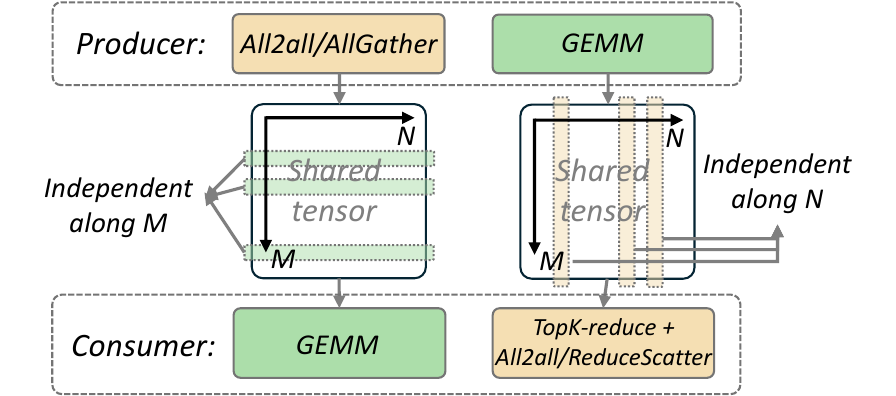}
	\caption{\label{fig:shared_tensor} The producer-consumer modeling of layer0 (left) and layer1 (right) of an MoE layer. The global size of the shared tensor is $(M\times topk, N)$ for both layer0 and layer1.}
\end{figure}

\subsubsection{How to decompose the shared tensor?}
Shared tensors, as the bridge between the producer operator and the consumer operator, is the key to enable overlapping. 
Notably, overlapping can occur only when the producer and consumer operate on independent data within the shared tensor, as illustrated in~\autoref{fig:shared_tensor}. Thus, we analyze the access pattern of operators on the shared tensor and decompose it along a specific dimension where data remain independent for the consumer operator.

For example, in the communication-computation pipeline in layer0, the consumer operator is a GEMM, with the shared tensor serving as its input matrix. In this case, tokens are independent with each other alongside the $M$ (token) dimension, allowing for decomposition of the shared tensor along $M$. 
However, since the computation of a GEMM tile involves multiplication and reduction along the token embedding dimension to produce the final outputs, decomposing the shared tensor along this dimension is not feasible.

As for the computation-communication pipeline in layer1, the consumer operator contains a top-K reduction, which reduces tokens along the $M$ dimension, leading to significant interdependencies between tokens along this dimension. Thus, the shared tensor can only be decomposed along the $N$ dimension where elements are independent.

\subsubsection{How to reschedule the decomposed shared tensor?}
At the finest granularity, the shared tensor can be split into individual rows or columns, enabling the consumer to begin computation as soon as a single row or column is received.
However, this level of granularity results in low computational efficiency, particularly in pipelines involving compute-intensive GEMMs, which are typically organized and processed in tiles to achieve high utilization.
Therefore, after decomposing shared tensors along specific dimensions, the resulting sub-tensors must be reorganized and rescheduled into tiles for computation. The rescheduling of shared tensors follows two principles:
\ding{172} \bluetext{Rescheduled sub-tensors should align with the original computation tile granularity for computational efficiency.}
\ding{173} The scheduling policy should prioritize portions of the producer that can be immediately used by the consumer, allowing the consumer to begin execution as early as possible.

\begin{figure}[t]
\centering

\includegraphics[width=0.98\columnwidth]{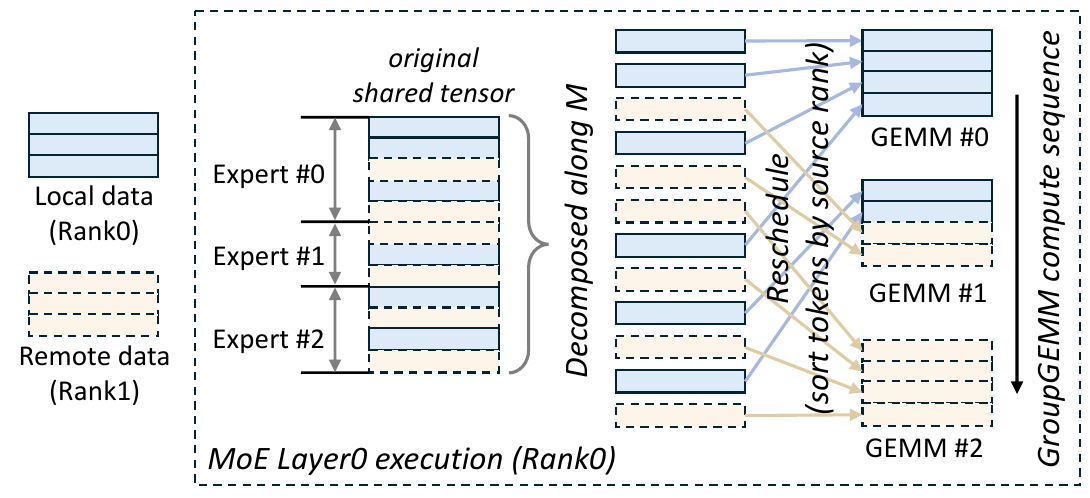}
	\caption{\label{fig:layer0} Decompose and reschedule the shared tensor in MoE layer0. In this illustration, three experts are located on Rank 0, each requiring both local and remote data for computation.}
\end{figure}

\begin{figure}[t]
\centering
	\includegraphics[width=0.9\columnwidth]{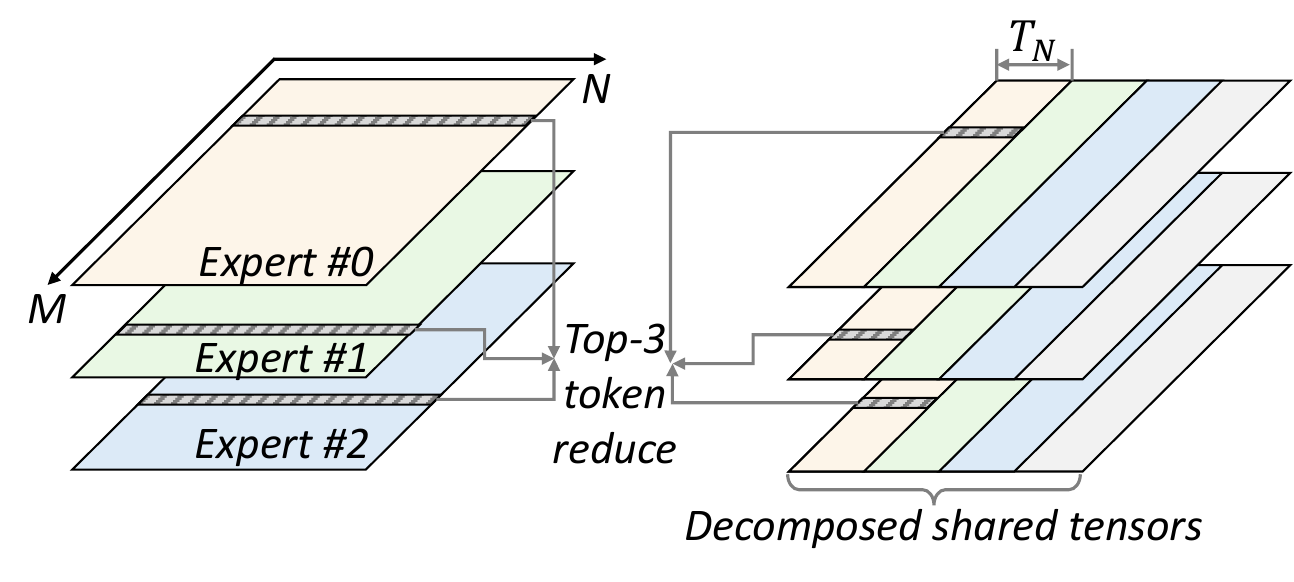}
	\caption{\label{fig:layer1} \bluetext{Rescheduled compute sequence for MoE layer1 ($E=3$ and $topk=3$). The execution order of the GroupGEMM is indicated by color (yellow $\rightarrow$ green $\rightarrow$ blue $\rightarrow$ grey). Here, $T_N$ denotes the tile size of a GroupGEMM along the $N$ dimension.}}
\end{figure}

\bluetext{\sysname{} leverages GroupGEMM to perform the computations for all experts on current rank.}
In the communication-computation pipeline (MoE layer0), the shared tensor, consumed by \bluetext{GroupGEMM}, is decomposed along the $M$ dimension. 
To enable early computation by the experts, tokens are sorted based on their source rank, as shown in~\autoref{fig:layer0}. The compute sequence of tiles \bluetext{in the GroupGEMM} is then designed to minimize dependency on remote data, with computation beginning from tiles containing local tokens while the transfer of other remote tokens proceeds concurrently.
In the computation-communication pipeline (MoE layer1), \bluetext{the shared tensor undergoes a top-k reduction after processing by the GroupGEMM of experts.}
As analyzed previously, the shared tensor is decomposed along the $N$ dimension. 
\bluetext{The tile computation sequence is adjusted (\autoref{fig:layer1}) to enable the consumer operator to start processing before expert computations are fully completed}.
Instead of computing each expert sequentially, \bluetext{GroupGEMM} operations are executed column-wise.
This approach allows the reduction and communicate operations to proceed as soon as the first $T_N$ columns of the shared tensors are computed.
Without rescheduling, tokens could only be reduced after all experts have completed their computations.

\subsection{Adaptive Workload Assignment}

With the decomposition and rescheduling of shared tensors, the pipelines in MoE can now achieve fine-grained overlap. To ensure effective latency hiding, the durations of fine-grained communication and computation must be closely aligned to minimize pipeline bubbles. Achieving this requires adaptive resource allocation for both computation and communication, tailored to specific tasks involved.

\begin{figure}[t]
\centering
	\includegraphics[width=0.86\columnwidth]{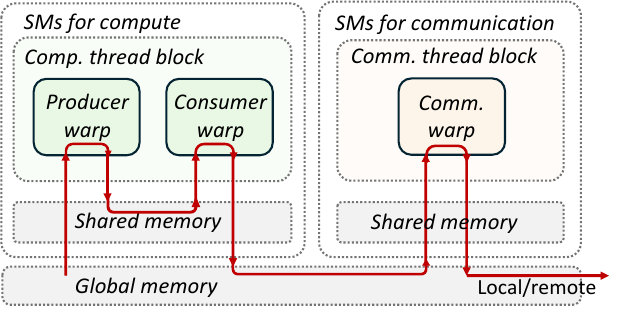}
	\caption{\label{fig:cta_division} Kernel design for the MoE layer1 on Hopper architecture. Each SM only accommodate one thread block. The red arrows indicates the route of data movement.}
\end{figure}


\subsubsection{Thread block specialization}
A straightforward approach to achieve communication-computation overlap in Mixture of Experts (MoE) is to encapsulate the entire pipeline within homogeneous thread blocks, integrating communication I/O into the prologue or epilogue of the computation (GEMM), a strategy referred to here as vertical fusion. 
Through vertical fusion, thread blocks execute concurrently, but the overlap occurs irregularly, leading to non-deterministic latencies of communication and computation, making it challenging to balance their durations for latency hiding.
Furthermore, token-level fine-grained I/O in MoE can significantly reduce the computational efficiency of the underlying kernels, particularly on advanced architectures such as Hopper. To address this, we implement thread block-level isolation between communication and computation workloads. This isolation enables precise control over hardware resource allocation for each workload, facilitating a balanced distribution between computation and communication that maximizes latency hiding.

\autoref{fig:cta_division} depicts the details of the thread block specialized kernel \bluetext{on Hopper}, with the critical data path highlighted in red. 
Due to the isolation between communication and computation, the GEMM thread blocks in \sysname{} utilize the same implementation as the default GEMM before fusion. In the scenario depicted in~\autoref{fig:cta_division}, where the GEMM is compiled using CUTLASS on the Hopper architecture, the GEMM execution is distributed across different warps. Specifically, the producer warp loads data from global memory into a shared memory buffer with the async TMA instructions, while the consumer warp initiates tensor core MMA operations~\cite{cutlass}.
The communication thread blocks subsequently read the results produced by the consumer warp from global memory. Following the top-K reduction, \bluetext{the warps} within the communication blocks either write tokens to the local global memory or transmit them to remote destinations. This thread block-specialized programming model is easily portable to other architectures, such as Ampere and Volta, requiring only a substitution of the respective compute thread block implementation.

\textbf{Hardware resource restriction.} 
The proposed thread block-specialized kernel is designed with the primary objective of minimizing data movement costs. However, this design must also contend with hardware resource limitations. For instance, it is theoretically feasible to integrate communication warps with computation warps within the same thread block to eliminate redundant global memory accesses.
However, the thread number restriction of warps constrict the communication operator to fully utilize the communication bandwidth.
From another perspective, the warps for communication also interfere with the computation warps within the same thread block.




\subsubsection{Adaptive thread block assignment}
\begin{figure}[t]
\centering
	\includegraphics[width=0.97\columnwidth]{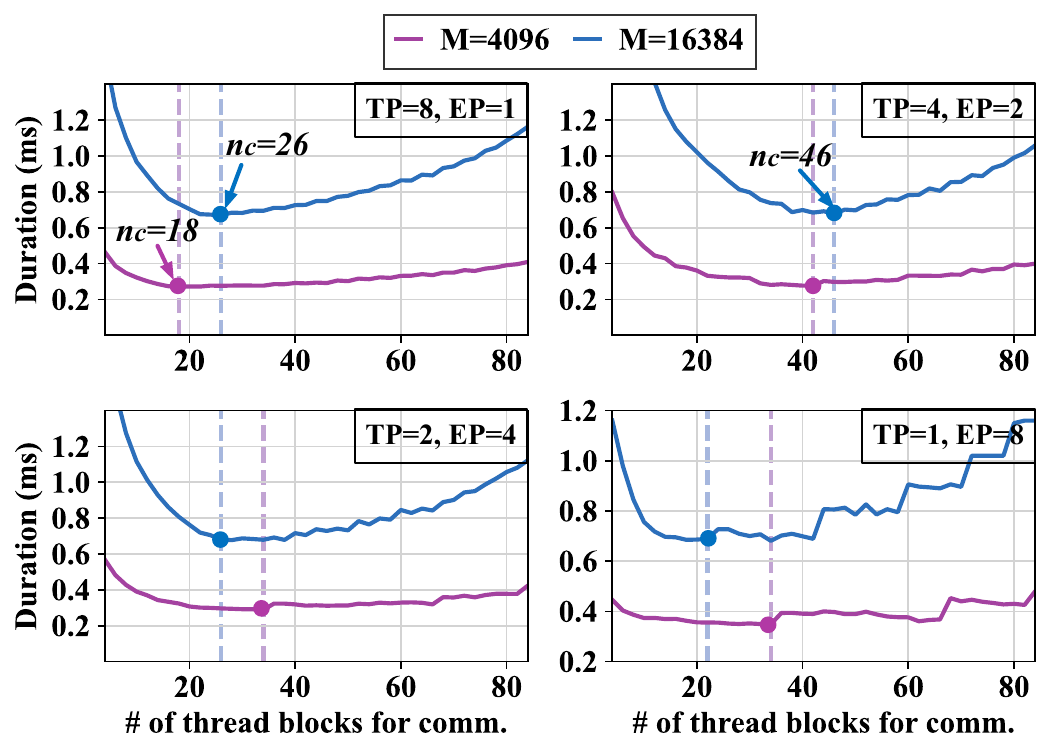}
	\caption{\label{fig:cta_num} Duration of the MoE layer1 kernel with varying number of thread blocks assigned for communication ($n_c$). The total number of thread blocks is identical to the number of SMs on Hopper(132). The figure shows four cases with different parallelisms.}
\end{figure}

Suppose that there are $n$ thread blocks for the fused kernel, within which $n_p$ blocks serve as producers in the pipeline and $n_c$ blocks serve as consumers. 
Identifying an optimal division point $n_p/n_c$ is crucial for maximizing overall efficiency.
We demonstrate that the optimal division point is influenced by the shape of input and specific model configurations in an MoE layer. To investigate this, we measure the duration of MoE layer1 across various input sequence lengths and parallelization strategies, as shown in~\autoref{fig:cta_num}.
It is observed that there exist an optimal division point under different configurations. 

When the input token length changes, although the data sizes processed by communication and computation operations both scale with input length, the scalability of the respective resource requirements differs.
Consequently, the optimal division point shifts with changes in input length.
For example, when $\textit{TP}=8$, the optimal $n_c$ changes from 18 to 26 when $M$ is changed from 4096 to 16384.
When the model configuration (parallel strategy) is modified, the optimal division point undergoes a significant alteration. For instance, when $\textit{TP}$ is adjusted from 8 to 4, the optimal $n_c$ is transformed from 26 to 46 with $M=16384$.

\sysname{}'s library comprises multiple pre-compiled kernels, each with a distinct division point. Prior to deployment, the optimal configuration for each setup is profiled and stored as metadata. During runtime, \sysname{} utilizes this metadata to select the optimal kernel for execution.


\section{Implementation}
\sysname{} consists of approximately 12k lines of C++ and CUDA code and 2k lines of Python.
\sysname{} provides a suite of user-friendly Python APIs and developers can seamlessly integrate the APIs into their frameworks. In production environment, \sysname{} has been implemented in Megatron-LM for large-scale MoE training. The source code will be available on GitHub.

\textbf{Optimized GEMM kernels for MoE.}
\sysname{} extensively utilizes the programming templates provided by CUTLASS to generate highly efficient GEMM kernels. Additionally, it incorporates various optimizations to minimize data movement overhead. For instance, in MoE layer 0, the row indices of the input matrix for GEMM operations must be accessed from global memory at each K iteration. By caching these row indices in registers, \sysname{} significantly reduces the global memory access cost.


\textbf{NVSHMEM as communication library.}
We employ NVSHMEM~\cite{nvshmem} within kernels to support fine-grained communication.
NVSHMEM is a communication library designed for NVIDIA GPUs. It creates a global address space for data that spans the memory of multiple GPUs and can be accessed with fine-grained GPU-initiated operations and CPU-initiated operations. 
Unlike NCCL~\cite{nccl}, which targets high-level communication operations, NVSHMEM offers a more composable, low-level API that facilitates finer data access granularity within kernels.

\section{Evaluation}

\begin{figure*}[t]
\centering
	\includegraphics[width=1.9\columnwidth]{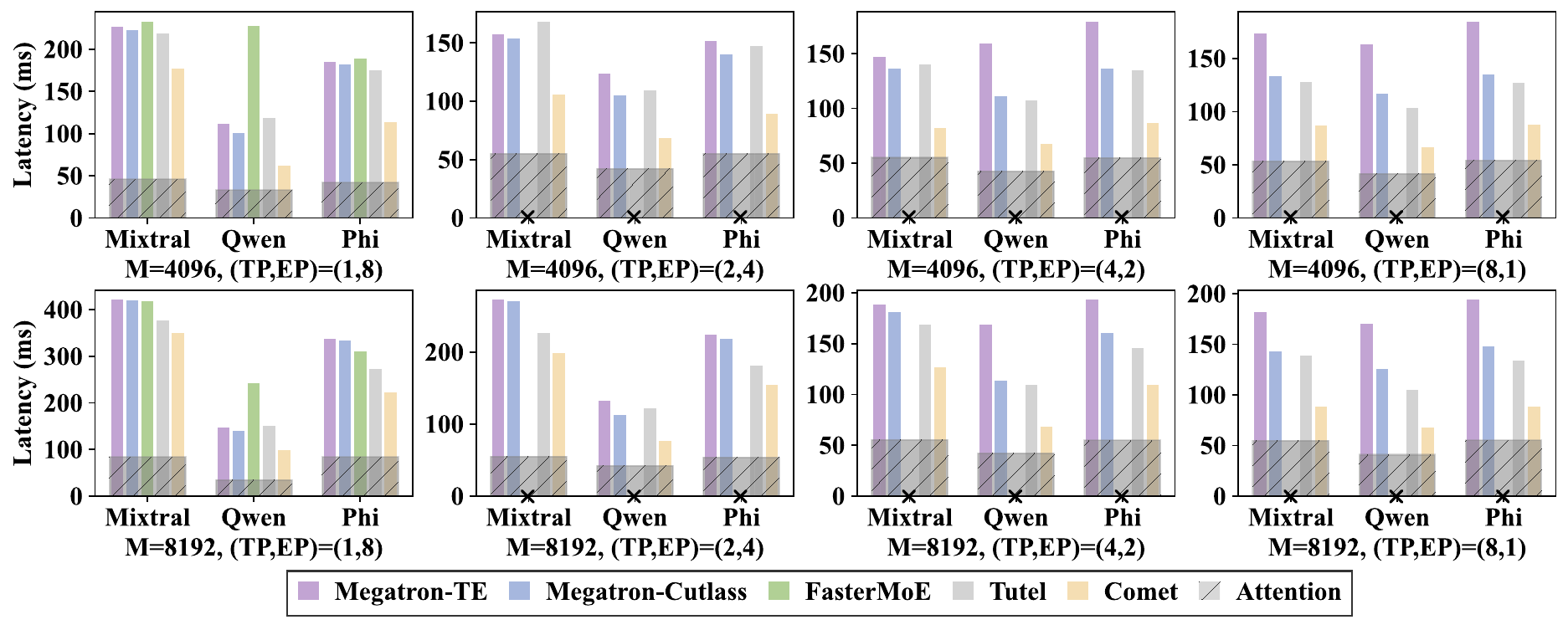}
	\caption{\label{fig:e2e} 
 End-to-end MoE model latency. 
 For the computation of MoE layers, the number of token on each device before permutation is $M\times W/\textit{TP}$.  
 The hatched region represents the identical duration of non-MoE (attention) layers in different mechanisms. Note that \textsc{FasterMoE} only supports expert parallelism for MoE layers.}
\end{figure*}

\begin{table}[]
\caption{\label{tab:configuration} Configuration of MoE models used in experiments. \browntext{The models are open-sourced on Hugging Face~\cite{huggingface}. The meaning of symbols are explained in~\autoref{tab:description}.} 
}
\footnotesize
\centering
\begin{tabular}{|l|c|c|c|c|c|}
\hline
                 & L & E & topk & N & K \\ \hline
Mixtral 8x7B     & 32         & 8           & 2    & 4096         & 14336             \\ \hline
Qwen2-MoE-2.7B & 24         & 64          & 4    & 2048         & 1408              \\ \hline
Phi-3.5-MoE      & 32         & 16          & 2    & 4096         & 6400              \\ \hline
\end{tabular}
\end{table}

\subsection{Experimental Setup}
\textbf{Testbed.} We evaluate \sysname{} on a server equipped with 8 Nvidia H800 GPUs (80 GB memory each). These GPUs are interconnected through NVLink. Our software environment includes CUDA 12.3, NVSHMEM 2.11, Pytorch 2.4.0 and Megatron-LM (git-hash 6dbe4c).

\textbf{Comparing targets.} 
We then compare \sysname{} with several baselines. All baselines are implemented on Megatron-LM, which is a widely adopted framework for high-performance model execution, integrating hybrid parallel strategies.

The baselines are: 
(a) \textbf{\textsc{Megatron-Cutlass}}: Megatron with MoE experts that are implemented through CUTLASS grouped GEMM~\cite{groupedgemm}.
(b) \textbf{\textsc{Megatron-TE}}: Megatron with experts that use transformer engine~\cite{te}. Transformer Engine is Nvidia's library for accelerating transformer models on NVIDIA GPUs.
(c) \textbf{\textsc{FasterMoE}}~\cite{fastmoe, fastermoe}: FasterMoE is an MoE system that customizes All-to-All communication to overlap the communication and computation operations of experts.
(d) \textbf{\textsc{Tutel}}~\cite{tutel}: Tutel delivers several optimization techniques for efficient and adaptive MoE, including adaptive parallelism, the 2-dimensional hierarchical All-to-All algorithm and fast encode/decode with sparse computation on GPU.


\subsection{Overall Performance}


We evaluate the end-to-end performance of \sysname{} in multiple large MoE models, including Mixtral 8x7B~\cite{jiang2024mixtral}, Qwen2-MoE~\cite{bai2023qwen} and Phi3.5-MoE~\cite{abdin2024phi}. The configurations of these models are shown in~\autoref{tab:configuration}.
The experiment is conducted with various input token lengths and diverse hybrid parallel strategies. The experimental details and results are shown in~\autoref{fig:e2e}. 
Note that when $\textit{TP}<W$, Megatron-LM enables data parallelism for non-MoE layers to improve overall throughput and the data parallel size is $W/\textit{TP}$. 
The computation of attention layers are identical with different mechanisms using Megatron-LM, and only the MoE layer is implemented differently with diverse mechanisms. 

As observed, the end-to-end latencies of the benchmarks are reduced by $34.1\%$, $42.6\%$, $44.4\%$ and $31.8\%$ with \sysname{} compared with \textsc{Megatron-Cutlass}, \textsc{Megatron-TE}, \textsc{FasterMoE} and \textsc{Tutel} respectively. 
The performance gain is more prominent with the identical attention computation apart.
\sysname{} outperforms other baselines in all configurations because it realizes sufficient overlapping and the scheduling inside high-performance fused kernels greatly reduce the the overhead at CPU side.

Besides, we can also observe that \textsc{Megatron-Cutlass} and \textsc{Megatron-TE} perform similar. This is because they are identical except from the implementation of \bluetext{GEMM/GroupGEMM. Neither of them supports overlapping, while \textsc{Megatron-TE} performs worse in some cases because of the overhead in transformer engine API calls}.
\textsc{Tutel} performs better than other baselines because it incorporates communication into experts' computation through delicate scheduling and adaptive parallelism. Although communication and computation is overlapped partially, when the number of experts is large (Qwen2), the advantage of \textsc{Tutel} diminishes because of the large scheduling overhead.
\textsc{FasterMoE} only supports expert parallelism ($\textit{EP}=W$) and it also does not perform well on Qwen2 because the experts are small in Qwen2 and the kernel invoking time for experts dominates the MoE layer. 

\subsection{Detailed Evaluation on a Single MoE Layer}
We then conduct an in-depth examination of a single MoE layer to perform a detailed analysis. 

\begin{figure}[t]
\centering
	\includegraphics[width=\columnwidth]{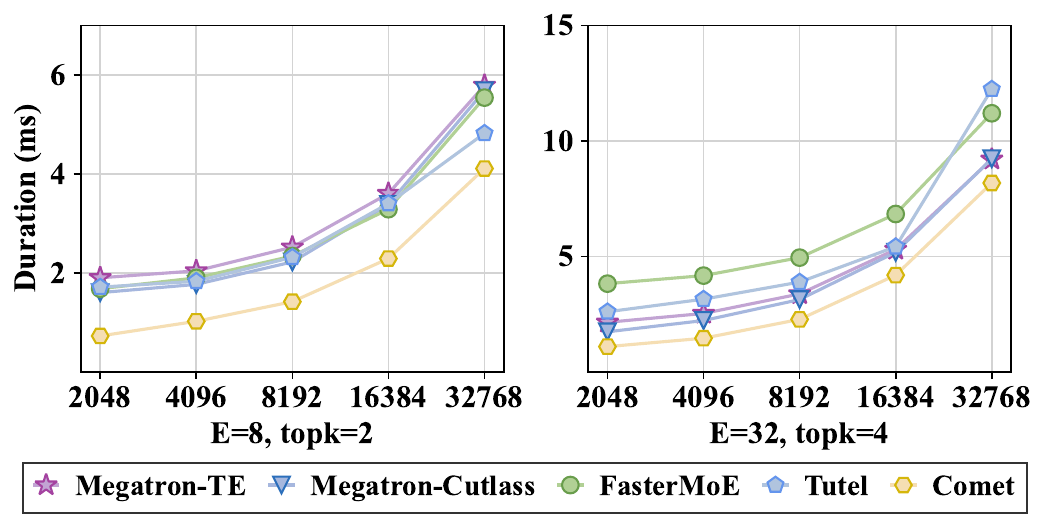}
	\caption{\label{fig:single_tp1} Single MoE layer duration with expert parallelism ($\textit{EP}=8$). The x-axis represents the total input token length $M$. Each device has $M/W$ tokens before token dispatching. The shape of experts are identical to that of Mixtral 8x7B.}
\end{figure}

\textbf{Handling varying input token lengths.}
The latency of a single MoE layer with varying input token lengths is shown in~\autoref{fig:single_tp1}. 
With the input token number varying, \sysname{} experiences a shorter duration compared with baselines and the improvement is stable.
\sysname{} achieves a $1.28\times$ to $2.37\times$ speedup compared with the baselines on average. 
It is noted that the advantage of \sysname{} is prominent especially when $M$ is small. This is because the scheduling time on the host side predominates the overall duration when $M$ is small and \sysname{} reduces such overhead \bluetext{through kernel scheduling within the fused kernel}. The scheduling overhead increases with $topk$ and $E$ for mechanisms with kernel-level scheduling (\textsc{FasterMoE} and \textsc{Tutel}) because the experts to manage become more complicated, inducing more kernels to be scheduled.

\begin{figure}[t]
\centering
	\includegraphics[width=\columnwidth]{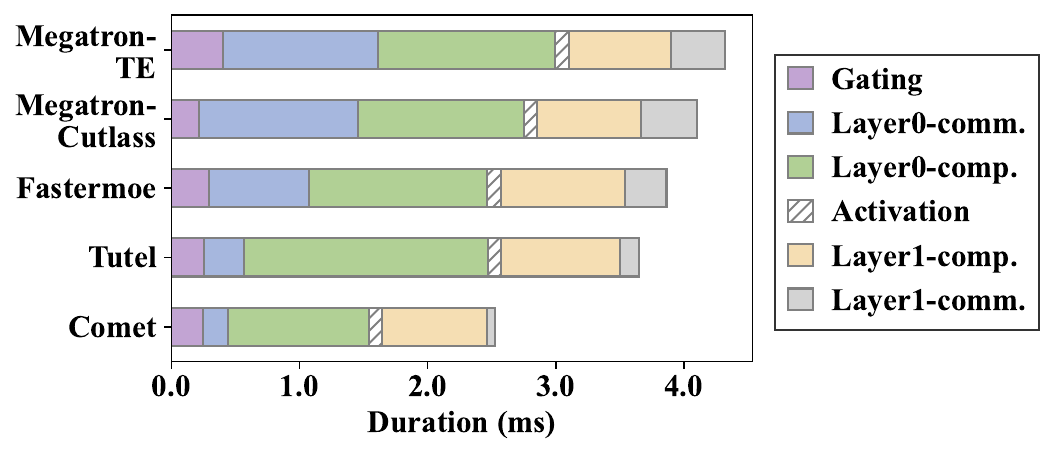}
	\caption{\label{fig:breakdown} Time breakdown of an MoE layer with expert parallelism. ($\textit{EP}=8, \textit{TP}=1, E=8, topk=2$ and $M=16384$).}
\end{figure}

\textbf{Time breakdown analysis of an MoE layer.}
The time breakdown of a specific MoE layer is shown in~\autoref{fig:breakdown}. 
Note that the communication part only consists of the GPU-to-GPU communication time, and the operations of token indexing, dispatching and combining on local device are regarded as the computation part.
As revealed, \textsc{Megatron-TE} and \textsc{Megatron-cutlass} experience no overlapping between communication and computation.
\textsc{FasterMoE} reduces the communication latency through customized Scatter and Gather operators, while the introduced local indexing extends the computation time.
\textsc{Tutel} reduces the communication overhead through the optimized all-to-all primitive design. However, its optimized all-to-all also exacerbates the burden of local computation.
\bluetext{\textsc{Megatron-TE} has no communication overlapped}. \sysname{} hides $86.5\%$ of communication latency on average and the computational efficiency of experts is not influenced, while \textsc{FasterMoE} and \textsc{Tutel} hide only $29.2\%$ and $68.6\%$ respectively. 


\begin{figure}[t]
\centering
	\includegraphics[width=0.7\columnwidth]{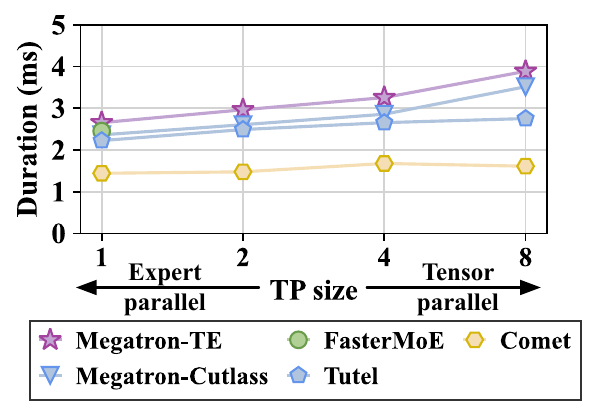}
	\caption{\label{fig:single_para} Single MoE layer duration under various parallelism strategies with $E=8, topk=2, M=8192, \textit{EP}\times \textit{TP}=8$.
 }
\end{figure}

\textbf{Parallelism within the MoE layer.}
Because of the introduction of expert parallelism, the parallel strategy within the MoE layer can be different from the model's overall parallel strategy.
\autoref{fig:single_para} shows the performance of methods applying diverse parallel strategies. Among all baselines, \textsc{FasterMoE} unfortunately does not support tensor parallelism.
For other baselines (\textsc{Megatron-TE}, \textsc{Megatron-Cutlass} and \textsc{Tutel}), the MoE layer latency increases when {\it TP} grows. This is because that tensor parallelism splits each expert onto multiple devices, triggering more fragmented small GEMMs for experts and resulting in a degradation of computational efficiency.
Nevertheless, \sysname{} maintains low latency in diverse parallelisms as \bluetext{the shared tensor is rescheduled to maintain computational efficiency} and the weight switching overhead is eliminated.

\subsection{Adaptiveness to Different Configurations}
We further inquire into the performance of \sysname{} when adapting different model configurations, runtime workloads and system environments.

\begin{figure}[t]
\centering
	\includegraphics[width=\columnwidth]{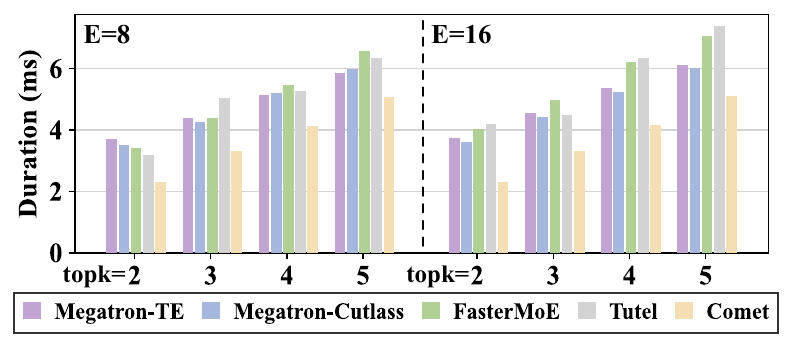}
	\caption{\label{fig:experts} Duration of a single MoE layer ($M=16384, \textit{EP}=8, \textit{TP}=1$) with various number of experts $E$ and $topk$.}
\end{figure}

\textbf{Performance with various MoE parameters.} We adjust the number of experts $E$ as well as $topk$ to evaluate the performance of \sysname{} in various MoE structures. The results are shown in~\autoref{fig:experts}.
With the increasing of $topk$, the duration of the MoE layer is increased because the computation amount at runtime is scaled up. 
\sysname{} consistently demonstrates superior performance across different values of $topk$ and $E$, yielding a speedup in the range of $1.16\times$ to $1.83\times$ compared to baseline implementations.

\begin{figure}[t]
\centering
	\includegraphics[width=0.9\columnwidth]{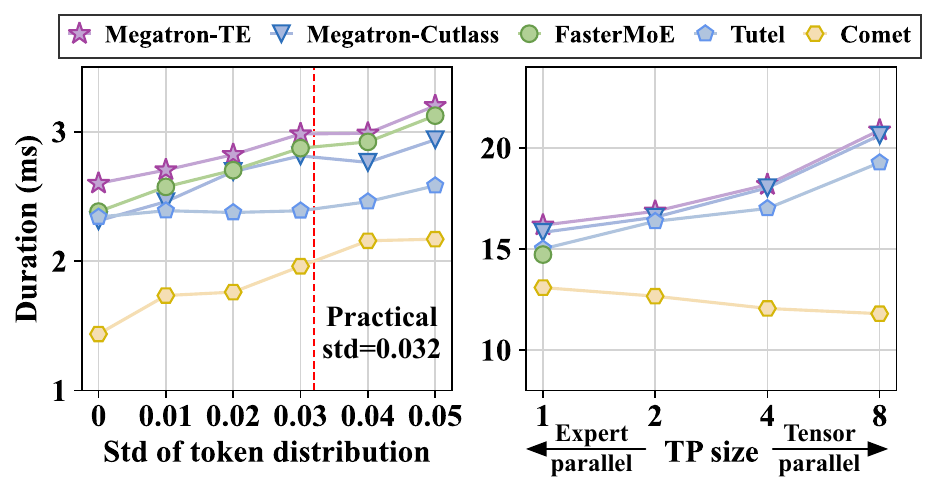}
	\caption{\label{fig:scaling} Performance of a MoE layer when scaling to different scenarios. \bluetext{Left: Duration with various token distribution with expert parallelism ($E=8, topk=2, M=8192, \textit{TP}=1, \textit{EP}=8$).} Right: Duration on a L20 Cluster with diverse parallelisms ($E=8, topk=4, M=8192, \textit{EP}\times \textit{TP}=8$). }
\end{figure}

\textbf{Performance with varying token distribution.}
When using expert parallelism, the number of tokens routed to different devices varies. We evaluate the performance of \sysname{} in scenarios with imbalanced token distribution. 
The standard deviation of the token distribution across different experts is denoted as $std$.
As shown in the left panel of~\autoref{fig:scaling}, 8192 tokens are distributed across various experts with differing distributions. When $std=0$, tokens are uniformly distributed and each expert receives $M\times topk / E = 2048$ tokens. 
At $std=0.05$, the least-loaded expert is assigned only a few hundred tokens. 
In a typical training job in production, the average $std$ is $0.032$.
When the load imbalance problem is exacerbated, the latency of the MoE layer in all systems is prolonged. \sysname{} consistently outperforms other MoE systems.

\textbf{Scaling to distinct clusters.}
We carry out the experiments on another distinct cluster with a different network environment. The cluster is equipped with 8 Nvidia L20 GPUs (46 GB memory) and the GPUs are connected via PCIe bridges. The GPU-to-GPU bandwidth is around 25 GB/s as tested, which is much lower than the H800 cluster. The experiments on the L20 cluster represents a bandwidth-limited environment.
As shown in the right panel of~\autoref{fig:scaling}, the average speedup of \sysname{} compared with other baselines is from $1.19\times$ to $1.46\times$. The results manifest the superiority of \sysname{} under different cluster environments.

\begin{table}[]
\caption{\label{tab:memory} Required device memory size for NVSHMEM.}
\footnotesize
\centering
\vspace{1mm}
\begin{tabular}{|c|c|c|c|}
\hline
Mem(MB) & Mixtral 8x7B & Qwen2-MoE & Phi3.5-MoE \\ \hline
M=4096     & 32           & 16        & 32         \\ \hline
M=8192     & 64           & 32        & 64         \\ \hline
\end{tabular}
 \vspace{-1mm}
\end{table}


\subsection{\browntext{Overhead Analysis}}

\sysname{} leverages NVSHMEM to allocate a shared memory buffer for communication on each device. The buffer size is dependent on the model configuration and equals to $MN$, where $M$ is the input sequence length and $N$ is the model hidden size. For datatype of BF16 or FP16, the allocated memory size is $2MN$. The communication buffer is global for the execution of the entire model, which means that it is shared across layers and experts. We list the device memory consumption of \sysname{} in~\autoref{tab:memory}, and it is negligible compared with the large device memory on current GPUs.

\section{Related Work}

With the successful application of MoE in large-scale distributed training and inference, there are plenty of works focusing on the system-level optimizations of reducing the communication overhead inherited in the MoE structure.

\paragraph{Communication optimization.}
To reduce the communication overhead in MoE execution, a straight-forward approach is to leverage efficient communication algorithms~\cite{all2all, shen2022se} for faster data transmission. Recent works~\cite{tutel, rajbhandari2022deepspeed, nie2022hetumoe} also propose the 2D-hierarchical all-to-all algorithm to better utilize intra-node bandwidth and accelerate MoE communication.
Some other works propose to reduce communication volume by data compression. For example, ScheMoE~\cite{schemoe} and Zhou et al.,~\cite{zhou2022accelerating} propose to apply data compression technologies to reduce the all-to-all communication volume while preserving the model convergence.

\paragraph{Computation-communication overlapping.}
The techniques of overlapping of computation and communication for dense models have been extensively employed in distributed training and inference~\cite{centauri, jangda2022breaking, song2023optimus, wang2022overlap, wang2023mgg, chang2024flux}.
For the MoE structure, recent studies also try to identify the pipelining opportunities for communication tasks of all-to-all operations and computing tasks of GEMMs.
FasterMoE~\cite{fastermoe} allows a pipeline degree of 2 to pipeline the expert computations and all-to-all communications. Tutel~\cite{tutel} enables a manually set degree of pipelining or a heuristic search under limited searching space, which may be sub-optimal. PipeMoE~\cite{pipemoe} 
and ScheMoE~\cite{schemoe} aim to schedule MoE operators to better utilize intra- and inter-connect bandwidths. These solutions realize overlapping through kernel-level scheduling 
and do not fully resolve the fine-grained data dependency in MoE.

\section{Conclusion}

In this paper, we propose \sysname{}, a MoE system that aims to achieve fine-grained communication and computation overlapping for MoE. \sysname{} features two key designs to achieve seamless overlapping without impact the computational efficiency: \bluetext{Shared tensor based dependency resolving that enables} fine-grained overlapping, while eliminating the bottleneck caused by fine-grained communication I/O; The workload assignment mechanism that promises precise and adaptive overlapping of operators, inducing maximal latency concealing.
\sysname{} achieves $1.96\times$ speedup in a single MoE layer and $1.71\times$ speedup in the end-to-end execution of MoE models, compared with existing literature.

\clearpage

\bibliographystyle{plainnat}
\bibliography{main}

\clearpage



\end{document}